\documentstyle[11pt,aaspp4]{article}

\begin{document}
\title{Quasars and Ultraluminous Infrared Galaxies: At the Limit?}
\author{K. K. McLeod}
\affil{Whitin Observatory, Wellesley College, 106 Central St., Wellesley, MA
02481; kmcleod@wellesley.edu}
\author{G. H. Rieke}
\affil{Steward Observatory, University of Arizona, Tucson, AZ 85721; grieke@as.arizona.edu}
\author{L. J. Storrie-Lombardi}
\affil{Observatories of the Carnegie Institution of Washington, 813
Santa Barbara Street, Pasadena, CA 91101; lisa@ociw.edu}

\begin{abstract}
We have detected the host galaxies of 16 nearby, radio-quiet quasars
using images  obtained with the Near-Infrared Camera and Multi-Object
Spectrometer (NICMOS).  We confirm that these luminous quasars tend to live in
luminous, early-type host galaxies, and we use the host-galaxy
magnitudes to refine the luminosity/host-mass limit inferred
from ground-based studies.  If quasars obey the relation
$M_{black hole}/M_{spheroid}\sim0.006$ found for massive dark objects
in nonactive galaxies, then our analysis implies that they radiate at
up to $\sim20\%$ of the Eddington rate.  An analogous analysis for
ultraluminous infrared galaxies shows them to accrete at up to similar
Eddington fractions, consistent with the hypothesis that some of them are
powered by embedded quasars. 

\end{abstract}


\section{Introduction}

Although active quasars constitute only a small fraction of galaxies
today, it appears that most large galaxies harbor central
massive dark objects (MDOs) with $M_{MDO}/M_{spheroid}\sim0.006$
(\cite{kr95}; \cite{fab97}; \cite{mag98}).  These MDOs are plausibly
the supermassive black holes required by the current paradigm for
active galactic nuclei.  This result has been predicted by studies of
quasar demographics that show quasar activity is more likely a
short-lived phenomenon in a majority of large galaxies than a
long-lived one in a small fraction of galaxies (\cite{sol82};
\cite{hae93}).  This conclusion is based on the assumptions that
quasars are found in massive spheroids and radiate at the Eddington rate.

Recently, quasar host-galaxy studies have added several pieces of
evidence in support of this picture.
First, near-infrared and HST imaging programs show that
luminous quasars do in fact reside mainly in luminous,
early-type hosts (\cite{mr95b};  \cite{hut95}; \cite{tay96}; \cite{m97};
\cite{bah97}; \cite{boy98}; \cite{mc98}).  Second,
there appears to be an upper bound to the quasar luminosity as a
function of host galaxy stellar mass (\cite{mr95a} and references
therein).  McLeod (1997) pointed out that if the quasar nuclei at this
``luminosity/host-mass limit'' are emitting at a significant fraction
of the Eddington limit, then the black holes must obey a relation
similar to the one for normal galaxies with MDOs.  This result
supports the notion that present day quiescent galaxies harbor ``dead
quasars" in their nuclei, a conclusion that has recently been
supported by McLure et al. (1998).

The nature of ultraluminous infrared galaxies (ULIRGs) has been of great
interest since their discovery by Rieke \& Low (1972), and
particularly since IRAS found them in substantial numbers (e.g.
\cite{san88}). It has been widely proposed that infrared galaxies with
luminosity of $\sim 10^{12}$ L$_\odot$ or greater derive their energy
predominantly from heavily dust-embedded AGNs (e.g. \cite{sm96} and
references therein).  If this model is correct, virtually all the
blue, ultraviolet, and soft x-ray energy from the quasar will be
absorbed by the dust and degraded to the far infrared, where the huge
luminosity will emerge unavoidably.  Comparison with the nuclear
luminosity/host-mass limit relation therefore can test the connection
between ULIRGs and quasars.

The quasar luminosity/host-mass limit was determined first from our
groundbased data (\cite{mr94a}b,1995ab and references therein), where
relatively poor resolution precluded a detailed study of the hosts.
We report here higher resolution imaging using HST's Near Infrared
Camera and MultiObject Spectrometer (NICMOS). We combine the results
with previous data to improve the definition of the relation. We then
use data from the literature to compare this relation with its
counterpart for ULIRGs.

\section{NICMOS Observations and Data Reduction}

To test the robustness of our luminosity/host-mass limit, we observed from our
``high-luminosity sample'' (\cite{mr94b}) all 10 quasars that had not
been previously observed with HST.  To this we added 6 luminous
quasars for which ground-based attempts to resolve a host galaxy had
failed.  All 16 objects are in the redshift range $0.13 < z < 0.40$
with an average $z=0.25$.

Each quasar was observed for a single orbit using the NIC2
MULTIACCUM mode in a four position dither pattern.  Because this mode
allows the observer to use intermediate readouts, we were able to
build up an image that was both deep and linear over the entire
quasar field.  At the end of each orbit, we used the same dither pattern to
observe a bright star near the quasar.  This star provided a
measurement of the point-spread function (PSF).  The
quasar and PSF star images were reduced using the NICRED package
provided by B. McLeod.  Full details will be provided in McLeod et al. 1999.

\section{Determining Host-galaxy Magnitudes}

We determined host-galaxy magnitudes using a 1-D analysis technique
that allowed us to compare with our ground-based results and provided
a graphical way to judge the goodness-of-fit of various galaxy
models. First, we  generated a 1-D radial intensity profile of each
quasar, each PSF star, and a ``combined PSF'' made from a
noise-weighted average of all of the PSF stars. The profiles extended
to surface brightness limit of $\rm m_H\approx23.2 mag/\arcsec^2$ and
excluded light from companions.  Second, we normalized
each PSF to have the same central intensity as the quasar.  Third, we
removed the nuclear contribution by subtracting the highest fraction
of the normalized PSFs for which the resulting intensity in the first
Airy minimum was non-negative.  On average, 72\% of the light in the
central peak was attributed to the nucleus, whereas the full range for the
sample was 65-90\%.  The combined PSF gave comparable results to each
quasar's own PSF.  Finally, we numerically integrated the resulting
profile to obtain a magnitude for the host.

\begin{figure}[ftbh]
\epsscale{0.7}
\plotone{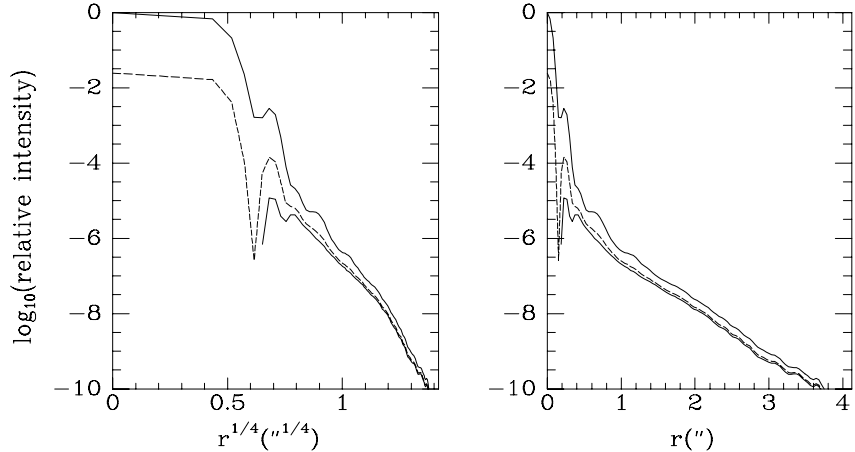}
\figcaption[sampleprof.eps]{
Sample radial profiles.  Profiles are of PG1352+183 plotted
v. $r^{1/4}$ (left) and $r$ (right).  Top solid line shows the
unsubtracted profile.  Dashed line shows the quasar profile after
subtracting the optimal fraction of the normalized PSF (see text).
Bottom solid line shows the quasar profile after subtracing 100\% of
the normalized PSF.  The central ringing is due to the Airy pattern.
\label{fig-sampleprof}}
\end{figure}

Figure \ref{fig-sampleprof} shows the 1-D profiles for one of the hosts
to illustrate the effect of subtracting different amounts of the
normalized PSF.  The host magnitudes derived by integrating under the
three profiles shown are H=14.0, 15.0, and 15.4 (top to bottom
respectively).  As seen in the Figure, most of the difference is in
the nuclear contribution; there is very little effect on most of the
galaxy.  We estimate from tests like these that our host galaxy
magnitudes generally have uncertainties of $\sim0.2$ mag from the PSF
subtraction.
Despite the order-of-magnitude difference in the spatial resolutions
of the ground-based and NICMOS data, we found excellent agreement in
the host magnitudes (within 0.2 mag) for 8 of the 10 quasars that we
had imaged previously. For the other two, the ground-based intensities were
apparently too high due to contamination from close companions.

\begin{figure}[ftbh]
\epsscale{0.7}
\plotone{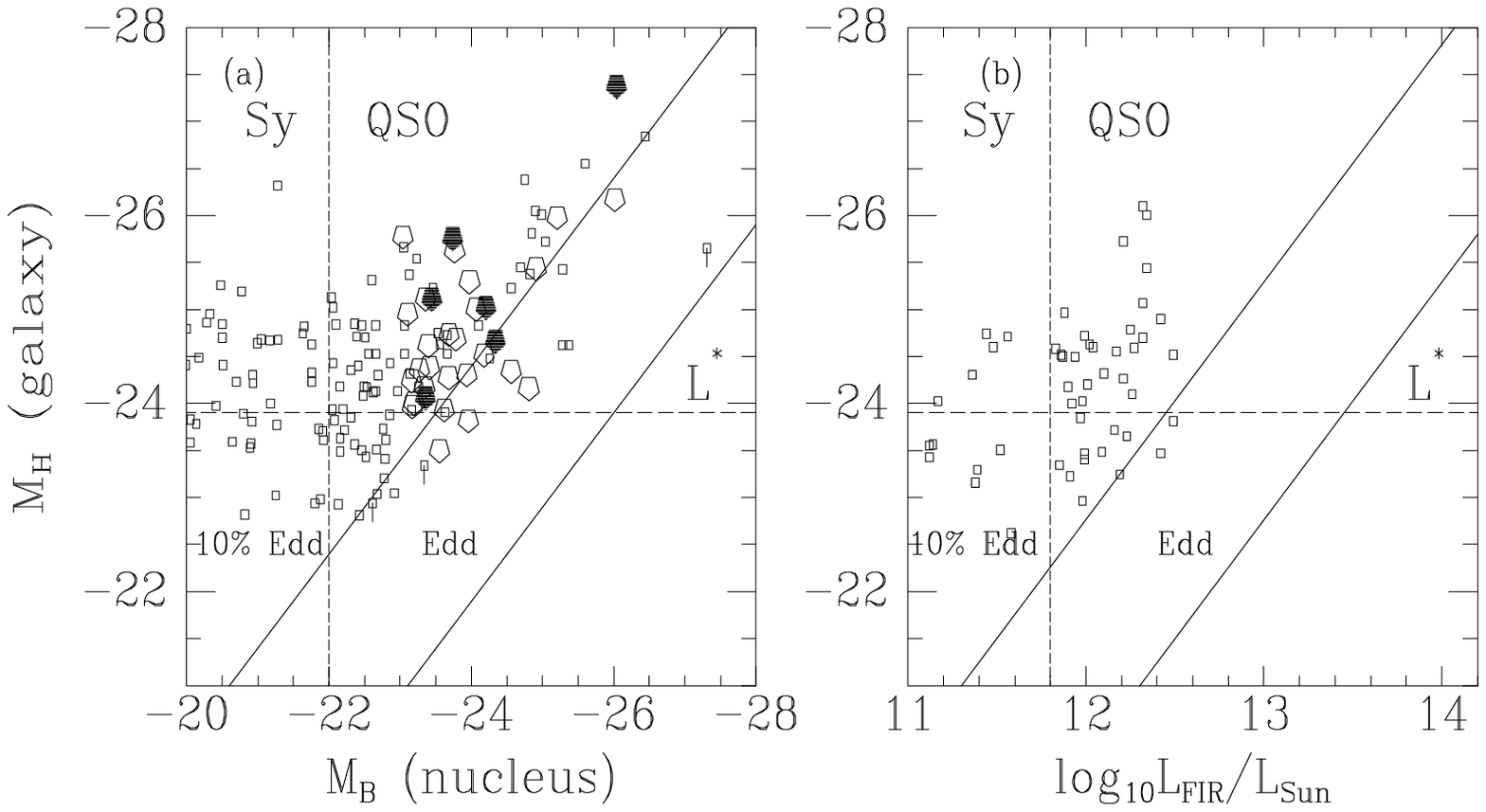}
\figcaption[allmags.eps]{
(a) Galaxy v. nuclear absolute magnitudes for QSOs.
Low-redshift QSOs and Seyferts shown as boxes are taken from
\cite{mr95a} and references therein, with additional new data from
\cite{ron96} (R-band imaging survey; three upper limits shown) and
\cite{hoo97} (WFPC2 imaging survey).  QSOs shown as open pentagons
constitute the high-luminosity sample from \cite{mr94b}, with host
magnitudes derived from either NICMOS images (\cite{m99})
or WFPC2 images (\cite{bah97}; \cite{mc98}).  Filled pentagons are the
other 6 QSOs from our NICMOS imaging.  Also shown are the
QSO/Sy boundary (dashed vertical line), position of an $L^*$ galaxy
(dashed horizontal line), and loci of Eddington and 10\% Eddington
luminosities for galaxies obeying the Magorrian et al. (1998) relation.
All values are rest-frame magnitudes with $\rm H_0=80km/s/Mpc,~q_0=0$.
(b) Galaxy v. far-infrared luminosity for the ULIRGs.  The axes cover
exactly the same range as in (a) for the bolometric correction
given in the text.
\label{fig-allmags}}
\end{figure}

The resulting magnitudes for all 16 quasars in our sample are plotted
in Figure \ref{fig-allmags}a, along with (i) WFPC2 host magnitudes from
the literature for the rest of the McLeod \& Rieke (1994b) sample;
(ii) our own and other ground-based data for other samples previously
shown in McLeod \& Rieke (1995a,b); and (iii) high-quality data
recently published for two additional samples.  All magnitudes have
been converted to rest-frame values assuming nuclear k-corrections and
colors from Cristiani \& Vio (1990); galaxy k-corrections and colors
appropriate for early-type galaxies, computed from a galaxy spectrum
provided by M. Rieke (and shown in Figure 1 of \cite{mr95b}); and $\rm
H_0=80km/s/Mpc,~q_0=0$.
The redshifts of the $\rm M_B<-22$ quasars on the plot
range from $0.06 < z < 0.8$ with an average $z\approx0.3$ (90\% have
$z<0.5$).  In addition to uncertainties due to PSF subtraction,
the host $\rm M_H$ values carry uncertainties due to the underlying
galaxy energy distribution.  For the galaxies observed in H, the
k-corrections themselves are $<0.1$ mag over this redshift range.  For the
galaxies observed in the visible, however, the uncertainties in the
k-corrections and colors can total several tenths of a magnitude.

  From the 1-D profiles, we also determined that approximately half of
these radio-quiet quasars have hosts that are better described by
deVaucouleurs laws than by exponentials, in agreement
with the other recent results mentioned above.  Full details of our 1-D
analysis, including the radial profiles, will be published along with
the images and a 2-D morphological analysis in McLeod et al. 1999.
However, the addition of the new data to Figure \ref{fig-allmags}a
allows us to take a new look at the luminosity/host-mass limit now.

\section{Results}

\subsection{The Eddington Limit for Quasars}

Figure \ref{fig-allmags}a allows us to test the combined assumptions
that all large galaxies contain black holes with the Magorrian et
al. (1998) mass fraction and that quasars radiate near the Eddington
limit.

For quasars near the luminosity/host-mass limit, the galaxy contribution to the
B-band light is negligible.  Thus, we can convert $\rm M_B$
to a quasar bolometric luminosity through a bolometric correction $\rm
BC\equiv\nu L_{\nu}(B)/L_{bol}$.  We adopt as a reference the
rest-frame value BC=12 (\cite{elv94}).

To estimate black hole masses, we adopt the average value of
$f\equiv M_{MDO}/M_{spheroid}\sim0.006$ from Magorrian et al. (1998).
We convert spheroid masses to galaxy absolute magnitudes assuming
$V - H$ = 3.0 for bulge stellar populations and a mass-to-light ratio
$\Upsilon_{\rm V}=7.2M_\odot/L_\odot$, which is an average for the 24 most luminous
galaxies ($\rm M_V \le -20$) in their sample.
Thus, both our bulge and black hole masses are traceable to Magorrian
et al. (1998), which should reduce systematic errors in the conversion
between them.

Combining the assumptions described above, the relation
between nuclear absolute B magnitude and bulge absolute H magnitude
is:

$$M_B=M_H-2.1-2.5[log_{10}(\epsilon)+log_{10}({\Upsilon_{\rm V}\over{7.2M_\odot/L_\odot}})+log_{10}({f\over{0.006}})-log_{10}({BC\over{12}})]$$

\noindent where $\epsilon\equiv L/L_{Edd}$.  The diagonal lines in Figure
\ref{fig-allmags}a show the positions of $\rm \epsilon=0.1~and~1.0$.
For our default values of $\rm \Upsilon_V$, f, and $\rm BC$, most quasars
fall within the $\rm\epsilon\approx0.20$ envelope.

A complementary analysis by Laor (1998) indicates that quasars like the ones in our samples do in fact follow the Magorrian et al. (1998) relation.
If this is true, our results imply that the quasars are radiating at
up to 20\% of the Eddington rate. These results are consistent
with the recent study of McLure et al. (1998) who carried out a
similar analysis to ours using visible data and
found most of their objects to be radiating at a few percent
Eddington.

\subsection{Bulge/Luminosity Relation for Ultraluminous Infrared Galaxies}

The discovery that classical quasars emit up to a significant fraction ($\sim
20\%$) of their Eddington luminosities suggests that ultraluminous galaxies
might emit at a similar level.  This hypothesis is
relatively easily tested because the integrated H-band fluxes of the
ULIRGs tend to be dominated by the bulge component of the galaxy (see e.g.
the imaging atlases of \cite{smi96} and \cite{mur96}; also argued by
\cite{sur97}). We therefore estimate the central black hole masses
from the integrated absolute H magnitudes, using  photometry
from McAlary et al. (1979), Carico et al. (1988),
Carico et al. (1990), Goldader et al. (1995), Smith et al. (1996),  and Murphy et al. (1996). For the latter reference, calibrated near infrared
measures are not available so we computed the bulge H magnitude from
m$_r$ and a standard color correction. In all cases, absolute
magnitudes are computed as in Murphy et al. (1996). Far infrared
luminosities are taken from the same references as the bulge
magnitudes. The results are shown in Figure \ref{fig-allmags}b, along
with an Eddington limit computed exactly analogously to the limit for
the quasars in Figure \ref{fig-allmags}a.

While they do not span a wide enough luminosity range to define clearly
a luminosity/host-mass relation, the $\sim 10^{12}$ L$_\odot$ ULIRGs
radiate at a rate nearly identical to that of quasars of
the same luminosity.  Most fall within an envelope of $\epsilon\lesssim0.1$,
and a detailed comparison of the $\epsilon$ distributions over the whole
QSO range indicates a factor of $\lesssim 2$ difference in the
average Eddington fraction.
This factor is likely within the uncertainties due to bolometric corrections,
assumptions about measuring the bulge luminosity in ULIRGs, and other
causes. It is conceivable that the two types of source have not just
very similar but identical behavior relative to the Eddington
limit. The close similarity supports the view that {\it a significant
portion of the most luminous ULIRGs derive much of their luminosity
from embedded AGNs}.  About 30\% of the ULIRGs above
Seyfert luminosity fall within a factor of two of $\epsilon = 0.1$
and are candidates to be dominated by embedded AGNs.

\subsection{Nature of ULIRGs}

The controversy regarding the nature of ULIRGs is fed because the
various indicators give contradictory results. Part of the difficulty
is that many indicators can show the presence of an AGN but do not
constrain its role in the energetics of the galaxy - an example is
high excitation optical emission lines, seen either directly or
scattered. Recently, Genzel et al. (1998) have used ISO spectroscopy
to argue that the majority (70-80\%) of these objects are dominated
energetically by star formation, both because the high excitation
infrared fine structure lines are weak and because the low excitation
lines imply adequate energy generation by starbursts. Rieke (1988)
concluded from the lack of hard xrays from this
class of object that most of them were powered by starbursts.
Although hard xrays could be blocked by very thick
accretion tori in some cases (\cite{sm96}), it is improbable that such
heavy columns would lie along our line of sight for all
cases. Surace et al. (1998) detect knots of star formation in HST
images, but they also detect putative nuclei and argue that the star
forming knots are not energetically important.
Veilleux, Sanders, \& Kim (1997) find evidence for
energetically important AGNs
in at least 20-30\% of the ULIRGs from the presence of near
infrared broad lines.  Lonsdale, Smith, \& Lonsdale (1995) find VLBI
radio sources at levels consistent with the theory that an AGN
dominates the luminosity in 55\% of ULIRGs.  On the other hand, in
followup VLBI observations of one such source, Arp 220, Smith et
al. (1998) show that the compact radio emission actually originates in
multiple radio supernovae.

All four indicators are consistent with $\sim$ 75\% of
the $\sim$ 10$^{12}$ L$_\odot$ ULIRGs being powered predominantly by
starbursts, with $\sim$ 25\% powered by AGNs. Although this general
agreement is encouraging, a number of individual galaxies yield contradictory
indications of their underlying energy source using differing
methods.

\section{Conclusions}

Our study of quasar host galaxies supports the hypothesis that all galaxy
spheroids contain black holes with $\sim0.6\%$ of the stellar
mass.  In QSOs, the black holes accrete at up to $\sim 20\%$ of the
Eddington rate.  In ULIRGs, the nature of the power source
remains unclear; however, if they are powered by embedded quasars then
they accrete at a similar rate.  The large Eddington fractions in both
kinds of objects imply a small duty cycle for activity over the
Hubble time.  Otherwise, the accretion process would produce
higher-mass black holes than we infer today.

At high redshift, quasars are very luminous but large galaxies might not
yet have been assembled.  Therefore, we expect that the
luminosity/host-mass limit must ultimately break down.  This may
indicate that rather than the stellar mass, it is the depth of the
large-scale potential well of the dark matter halo that is fundamentally
related to the mass of the supermassive black hole.

\acknowledgments Support for this work was provided by NASA through
grant number GO-07421.01-96A from the Space Telescope Science Institute, which
is operated by the Association of Universities for Research in
Astronomy, Inc., under NASA contract NAS5-26555.  We also gratefully
acknowledge support from the Keck Northeast Astronomy Consortium.  We
especially thank the NICMOS team for putting up a great instrument,
Brian McLeod for putting together the NICRED reduction package, and
Erin Condy for assistance with the data reduction.  We are grateful to
the referee for a concise, constructive report that improved the
quality of the manuscript.

\end{document}